\documentclass{elsarticle}
\input epsf
\usepackage{amsmath,amssymb}
\usepackage{epsfig}
\usepackage{graphicx}
\journal{New Astronomy}

\def\mathrelfun#1#2{\lower3.6pt\vbox{\baselineskip0pt\lineskip.9pt
  \ialign{$\mathsurround=0pt#1\hfil##\hfil$\crcr#2\crcr\sim\crcr}}}

\def\ln {{\rm ln}}

\def\Mnu{M_\nu}
\newcommand{\LCDM}{$\Lambda$CDM}
\newcommand{\nnu}{$0\nu \beta \beta$}
\newcommand{\Thalf}{T_{1/2}^{0\nu}}

\begin{document}

\title{Coupling between cold dark matter and dark energy from neutrino
mass experiments}

\author{J.R. Kristiansen$^1$, G. La Vacca$^{2,3}$, L.P.L. Colombo$^4$,
R. Mainini$^{1,5}$,\\ S. A. Bonometto$^{2,3}$
\\ $^1$IInstitute of Theoretical Astrophysics, University of Oslo, Box 1029, 0315 Oslo, Norway
\\ $^2$Physics Dept.~G.$\, $Occhialini, Milano--Bicocca
University, Piazza della Scienza 3,\\ 20126 Milano, Italy
\\
$^3$I.N.F.N., Sez.~Milano--Bicocca, Piazza della Scienza 3,
20126 Milano, Italy
\\ $^4$Physics Dept.,
University of Southern California, Los Angeles, CA 90089-0484
\\ $^5$I.N.A.F.,
Roma Observatory, Via Frascati 33, 00040 Monteporzio Catone, Italy\\
E-mail: j.r.kristiansen@astro.uio.no, lavacca@mib.infn.it}


\begin{abstract}
We consider cosmological models with dynamical dark energy (dDE)
coupled to cold dark matter (CDM), while simultaneously allowing
neutrinos to be massive. Using a MCMC approach, we compare these
models with a wide range of cosmological data sets. We find a strong
correlation between this coupling strength and the neutrino mass. This
correlation persists when BAO data are included in the
analysis. We add then priors on $\nu$ mass from particle
experiments. The claimed detection of $\nu$ mass from the
Heidelberg-Moscow neutrinoless double--$\beta$ decay experiment would
imply a $7$--$8 \sigma$ detection of CDM-DE coupling. Similarly, the
detection of $\nu$ mass from coming KATRIN tritium $\beta$ decay
experiment will imply a safe detection of a coupling in the dark
sector. Previous attempts to accommodate cosmic phenomenology
with such possible $\nu$ mass data made recourse to a $w < -1$ eoS.
We compare such an option with the coupling option and find that the
latter allows a drastic improvement.

\end{abstract}

\maketitle

\section{Introduction}
Today we have a standard model of cosmology, the so-called \LCDM{}
model, providing an excellent fit to all cosmological data. This model
tells us that dark energy (DE) and cold dark matter (CDM) account of
$\sim 75\, \%$ and $20 \, \%$ of the present cosmic energy budget,
respectively \cite{dunkley:2008, komatsu:2008}. It is then quite
embarrassing that we fail to understand the nature of both DE and
CDM. Furthermore, if DE is a smooth cosmological {\it fluid} with an
equation of state $w=-1$, our
model is troubled by
two fundamental questions related to its relative and absolute
density, the {\it coincidence} and {\it fine tuning} problems: Why did DE
start to dominate just when structures had time to form? Why is its
density $\sim$120 orders of magnitude smaller than the (na\"ively)
expected quantum vacuum density?

One way to circumvent the coincidence problem amounts to introducing a
coupling between CDM and dynamical DE (dDE) \cite{wetterich:1995,
amendola:1999}. The energy transfer from CDM then allows DE to
comprise a significant fraction of the cosmic energy budget 
over a large part of the cosmic history. 
Unfortunately, switching on a coupling apparently worsens the fit of
cosmological data.

More recently, however, it has been noted that neutrino masses
($m_\nu$) and a CDM-dDE coupling affect cosmological observables in
opposite ways \cite{lavacca:2008, lavacca:2009}. 
However, not only our ignorance about $m_\nu$ softens the
constraints on the coupling, but a much more puzzling effect
arises: models with $m_\nu \neq 0$ and coupling appear (slightly)
favored in respect to $m_\nu \sim 0$ uncoupled models.

Neutrinos are abundant in the Universe, second only to photons when it
comes to number density; $\nu$ oscillation experiments tell us that
they are massive \cite{messier:2006} and measure the mass split
between $\nu$--mass eigenstates, so that the largest split ($\sim
0.05$eV from atmospheric $\nu$'s) is a lower limit on the heaviest
$\nu$ mass.

Particle experiments have placed various upper limits on the absolute
$\nu$ mass scale. The Mainz and Troitsk experiments, measuring the
end--point of the electron energy distribution in tritium $\beta$
decay, gave a 95\% C.L. limit $m_{\beta} < 2.0$eV \cite{otten:2008}. A
further controversial detection of an absolute $\nu$ mass came from
the~Hei\-del\-berg-Moscow (HM) experiment; on the basis of its
outputs, a part of the experiment team claims a $3 \sigma$ lower
limit $m_{\beta \beta} = (0.2-0.6)$eV \cite{klapdor:2004,
klapdor:2005} (the mass measured by neutrinoless double $\beta$ decay
(\nnu{}) experiments, $m_{\beta \beta}$, is however a different
combination of mass eigenvalues than the mass $m_{\beta}$, from
tritium $\beta$ decay).

In 2011, the experiment KATRIN \cite{katrin:2004}, also studying
tritium $\beta$ decay, is expected to start taking data. With a
prospected sensitivity of $\sigma_{m_{\beta}^{2}} \approx 0.025
\textrm{eV}^2$, it should be able to detect $m_\nu$ values in the
range of the $m_{\beta\beta}$ claim.

At present, the best upper limits on the $\nu$ mass scale come from
cosmology, yielding $\Sigma m_\nu \equiv \Mnu \lesssim 0.2$eV
\cite{goobar:2006, seljak:2006} and $\Mnu \lesssim 1.5$eV (at 95\%
C.L.), depending on data sets and  cosmological models
\cite{goobar:2006, seljak:2006, lesgourgues:2006, kristiansen:2007,
zunckel:2006, dunkley:2008, komatsu:2008, reid:2009}.

All this is true if the possibility of a dDE--CDM coupling is
neglected. 

In this paper, similarly to what is done in \cite{lavacca:2009}, we
will allow for such a coupling, reporting for the first time results
when baryonic acoustic oscillation (BAO) data are taken into account.
Then we consider the $\nu$ mass limits set by a part of the HM
collaboration (KKDC claim, herafter), and impose this as a prior on
$M_\nu$. Finally we shall investigate how a $m_\beta$ detection from
KATRIN would affect limits on coupling.

The possibility that forthcoming neutrino experiments yield mass
values in apparent conflict with cosmic data had been considered by
various authors. The most promising option, perhaps, had been
discussed by \cite{anne,KE}, who found that allowing $w$, the state
parameter of DE, to take values $< -1$ eased such constrast. Here we
compare this option with the coupling option and find that the latter
one leads to a drastically better agreement between terrestrial and
cosmological measures.

In the following section we discuss our cosmological model and outline
the observable effects of the CDM-dDE coupling and massive $\nu$'s.
In Section \ref{sec:nu} we discuss the experimental bounds on the
absolute $\nu$ mass scale. The data and methods used are presented in
Section \ref{sec:data}, while Section \ref{sec:results} is devoted to
reporting and discussing our results. In Section \ref{sec:summary} we
summarize our findings and conclude.

\section{Cosmological model} \label{sec:model}

Our models differ from the standard $\Lambda$CDM in three different
aspects: (i) {DE is a self--interacting scalar field $\phi$ rather
than a cosmological constant $\Lambda$.} (ii) {A linear dDE--CDM coupling is
allowed.} (iii) {We allow $\nu$'s to be massive.}

We shall consider the Ratra-Peebles (RP) potential \cite{ratra:1987}
and a SUGRA self--interaction potentials \cite{brax:2000}, reading
\begin{equation}
V(\phi) = \Lambda^{\alpha+4}/\phi^{\alpha}
~,~~~ V(\phi) = (\Lambda^{\alpha+4}/\phi^\alpha) \exp(4\pi
\phi^2/m_p^2) ~,
\end {equation}
respectively; they allow tracker solutions for any $\alpha >
0$. For both potentials, once $\alpha$ and $\Lambda$ are assigned, the
DE density parameter $\Omega_{o,DE}$ is uniquely defined. In our
fitting procedure, however, we use $\Omega_{o,DE}$ and $\lambda = \log
(\Lambda/{\rm GeV})$ as free parameters.

Limits on these models {\it without} coupling between DE and CDM have
been studied in \cite{lavacca:2009b}. For most cosmological
parameters, WMAP5 results lead to a slight narrowing of error bars, in
respect to WMAP3. In the case of $\lambda$, however,
\cite{lavacca:2009b} find a significant shift downward of the
2--$\sigma$ upper limit on the energy scale $\Lambda$. In the SUGRA
case, in particular, only $\lambda \lesssim -3.5$ is allowed. Such small
values are well below the range motivated by particle
physics. Therefore the physical appeal of the SUGRA potential is
spoiled.

Following the procedure in refs. \cite{lavacca:2008, lavacca:2009} we
assume a linear coupling between the DE and CDM energy components,
which leads to the coupled equations
\begin{equation}
\ddot \phi + 2\frac{\dot a}{a} \dot \phi + a^2 V'(\phi) = Ca^2 \rho_c
\end{equation}
\begin{equation}
\dot \rho_c + 3 \frac{\dot a}{a} \rho_c = -C \dot \phi \rho_c~;
\end{equation}
here $a$ is the scale factor, $\rho_c$ is CDM density, $C$
denotes the coupling strength. We then define the dimensionless
coupling parameter
\begin{equation}
\beta \equiv \sqrt{3/16\pi} \ m_p C, 
\end{equation}
used as a free parameter for our cosmological models. For a more
thorough discussion of the effects of coupling on dDE and CDM
evolutions, see, {\it e.g.}, \cite{lavacca:2008}.

Let us however outline that, when the $\beta$ degree of freedom
is opened, $\Lambda$ values as large as 30$\, $GeV become allowed, at
the 1--$\sigma$ level, while, at the 2--$\sigma$ level, no significant
constraint on the energy scale $\Lambda$ remains. Even for the RP
potential, for which a limit $\lambda \lesssim -8.5$ held, in the absence
of coupling, values $\lambda \sim -2$ become allowed.

The effects of massive neutrinos in cosmology have been studied
thoroughly for many years, and we refer to \cite{lesgourgues:2006} for
an extensive review of the topic. Cosmological observations are mostly
sensitive to the sum of $\nu$--masses, $\Mnu$, related to the $\nu$
density parameter by the relation $\Omega_\nu h^2 = {M_\nu}/{93.14
\textrm{eV}}$.

If $\Mnu \lesssim 4.5\, $eV, most $\nu$'s were still relativistic at
matter-radiation equality and act then as radiation, so postponing the
equality compared to a model 
where all the DM is cold.
When keeping the
total DM density constant, increasing $\Omega_\nu$ shifts the peaks of
the anisotropy spectrum $C_\ell$ to lower $\ell$ and boosts their
heights.

When it comes to the matter power spectrum $P(k)$, the main effect is
$\nu$ free-streaming from small scale fluctuations, damping $P(k)$ for
large wavenumbers~$k$.

The combined effects on $C_\ell$ and $P(k)$ spectra, when compared
with observations, lead then to the limits outlined above
\cite{goobar:2006, seljak:2006, lesgourgues:2006, kristiansen:2007,
zunckel:2006, dunkley:2008, komatsu:2008}. These effects, however, as
outlined in \cite{lavacca:2008, lavacca:2009}, are almost opposite to
those of dDE--CDM coupling, for both $C_\ell$ and $P(k)$.  A
Fisher--matrix analysis then shows a strong degeneracy between the
$\beta$ and $\Mnu$ parameters. Pinning down one of the parameters by
some other means clearly results in improved limits on the other
parameter.

\section{Neutrino mass bounds from earth based experiments} \label{sec:nu}

Two different ways are being followed to measure the absolute scale of
$\nu$ masses: \nnu{} and tritium $\beta$
decay experiments.

The \nnu ~process can only occur if $\nu$'s are massive Majorana
spinors, {\it i.e.} they coincide with their own antiparticles. From
the measurement of $T^{0\nu}_{1/2}$ (\nnu \ half life) one can then
deduce an effective mass $ m_{\beta \beta} \equiv \sum_i U_{ei}^2 m_i
$, being
\begin{equation}
m_{\beta \beta}^2 = \frac{m_e^2}{C_{mm}T^{0\nu}_{1/2}}.
\end{equation} 
Here $U_{ei}$ is the PMNS $\nu$ mixing matrix, $m_i$ are mass
eigenvalues, $m_e$ is the electron mass. $C_{mm}$,
the
nuclear matrix element relevant for the nuclide considered, is the
main problem with \nnu, because of its large theoretical uncertainty.

The most sensitive limits, up to now, were derived from detectors
enriched in $^{76}$Ge. Using such nuclide, the Heidelberg--Moscow (HM)
\cite{baudis} and the IGEX \cite{igex} gave the limits
$T^{0\nu}_{1/2} > 1.9 \times 10^{25} y$ and $T^{0\nu}_{1/2} > 1.6
\times 10^{25} y$, respectively. However, a part of the HM
collaboration published results claiming a more than $5 \sigma$
detection of a signal, giving an effective neutrino mass of $m_{\beta
\beta} = (0.2-0.6)$eV ($3 \sigma$ limits) \cite{klapdor:2004,
klapdor:2005} (which we refer to as the KKDC claim).

In view of nuclear matrix uncertainties, this claim is not in
contradiction with another \nnu{} experiment, CUORICINO
\cite{arnaboldi:2008}, based on $^{130}$Te, placing an upper limit of
$m_{\beta \beta} < \{ 0.19 - 0.68 \}$eV.

Both $^{130}$Te and $^{76}$Ge double--$\beta$ decays are however still
being investigated through the CUORE and the GERDA experiments,
respectively. The CUORE experiment \cite{arditi}
is running in the Laboratori Nazionali del Gran Sasso (LNGS), where
the GERDA experiment \cite{gerda:2008} is also being placed.

Tritium $\beta$ decay is a different road to the absolute neutrino
mass scale, where one accurately measures the energy distribution of
the outgoing electron, and uses this to infer an effective electron
neutrino mass, $m_\beta^2 = \sum_i \left| U_{ei} \right| m_i^2$.

Currently, the best limits on $m_\beta$ from tritium $\beta$ decay
come from the Mainz and Troitsk experiments, yielding an upper limit
of $m_\beta < 2.0\, $eV (95\% C. L.). The much more sensitive KATRIN
experiment \cite{katrin:2004} is scheduled to start taking data this
year. When completed, KATRIN is expected to reach a sensitivity of
$\sigma_{m_\beta^2} \approx 0.025\, \textrm{eV}^2$, and thus be able to
confirm the KKDC claim.

\section{Data and methods} \label{sec:data}

We use a modified version of the public program CAMB
\cite{Lewis:1999bs} to calculate CMB and matter power spectra. A
modified version of the public MCMC~engi\-ne CosmoMC \cite{lewis:2002}
is used to generate confidence limits on the parameters: \{$\omega_b$,
$\omega_{c}$, $\theta$, $\tau$, $n_s$, $\ln \ 10^{10} A_s$, $\log_{10}
\Lambda$, $\beta$, $M_\nu$\}. Here $\omega_{b}, \omega_{c}$ are the
reduced density parameters of baryons, CDM; $\theta$ is the ratio of
the sound horizon to the angular diameter distance, $\tau$ is the
optical depth, $n_s$ and $A_s$ are the primordial scalar spectral
index and amplitude (at $k=0.05 \textrm{Mpc}^{-1}$). We
marginalize over SZ amplitude. The
parameters above are given flat priors, unless otherwise is explicitly stated.

We mostly use two different combinations of cosmological data sets;
WMAP5 data only, and WMAP5 plus other cosmological data (referred to
as WMAP5++). For the sake of comparison, in a specific case, we
also consider the option of omitting BAO constraints, as specified
below.

We then apply priors on $\nu$ mass according to the KKDC claim and the
prospected KATRIN results. Data sets and priors used are described in
the following. The MCMC chains are run on the Titan cluster at Oslo
University.

\subsection{Cosmological data}
Firstly, we tested our models using only the five year data from the
WMAP measurements of the CMB radiation (WMAP5) \cite{nolta:2008,
komatsu:2008, dunkley:2008}. The WMAP5 data are analyzed with the
Fortran 90 likelihood code provided with the data release.

In WMAP5++ we then included the galaxy power spectrum from the 2dF
survey \cite{cole:2005}, SNIa data from the SNLS survey
\cite{astier:2006}, and added gaussian priors on the Hubble parameter
of $h=0.72 \pm 0.08$ from the HST key project \cite{freedman:2001},
the physical baryon density, $\omega_b = 0.022 \pm 0.002$
\cite{burles:2001, cyburt:2004, serpico:2004} inferred from the $^4$He
abundance after Big Bang nucleosynthesis, and BAO data from
\cite{percival:2009}.

The BAO analysis in \cite{percival:2009} includes both the SDSS DR7
data and data from 2dF. Angular distance measurements are compared at
$z=0.2$ and $z=0.35$, by using the distance measure $d(z) \equiv
{r_s(z_d)}/{D_V(z)}$. Here $r_s(z_d)$ is the comoving sound horizon at
the baryon drag epoch (see \cite{percival:2009, eisenstein:1998}) and
$D_V(z) \equiv \left[ (1+z)^2 D_A^2 z / H(z) \right]^{1/3}$, where
$D_A$ is the angular diameter distance.  This provides information on
the late expansion of the Universe, which is important to constrain DE
nature. Following \cite{percival:2007, percival:2009}, from the BAO
analysis, the inferred likelihood $\mathcal L$ of a model can be
estimated by $-2 \ \ln \ \mathcal L \propto \mathbf{X^{-1}C^{-1}X}$,
where
\begin{equation}
{\bf X} = \left| 
\begin{matrix} d(0.2) - 0.1905 
\cr d(0.35) - 0.1097 
\end{matrix}
\right| ~,~~~
{\bf C}^{-1} = \left| 
\begin{matrix} 30124 & -17227 
\cr -17227 & 86977 
\end{matrix}
\right|~
\end{equation}

\subsection{KKDC and KATRIN priors}

In order to include KKDC priors on $\nu$ mass, we use the procedure in
refs.~\cite{macorra:2006, fogli:2006}. To account for the dispersion
in nuclear matrix ($C_{mm}$) estimates, we follow \cite{fogli:2006},
where upper and lower extremes from a compilation of {\it reliable}
theoretical estimates are used, defining a $3\sigma$ uncertainty width
of $C_{mm}$. Combined with the uncertainty in $\Thalf$, this results
in a KKDC prior of $\log_{10}(m_{\beta \beta}/\textrm{eV}) = -0.23 \pm
0.07$ (at $1 \sigma$). 
In this mass range the $\nu$ mass
eigenvalues are almost degenerate, so that we can use that $M_{\nu} =
3m_{\beta \beta}$ without any loss of accuracy.

As far as KATRIN is concerned, we use their expected uncertainty
$\sigma_{m^2_{\beta}} = 0.025\, ${eV}$^2$ for a Gaussian distribution
around a best-fit value $m^2_{\beta}$, as in refs. \cite{host:2007,
KE, hannestad:2007}. We then assume a fiducial value
$m_\beta = 0.3\, $eV ($M_\nu = 0.9\, $eV), as a compromise between
cosmological upper limits and KKDC lower limit.

KKDC and KATRIN priors are imposed in the post processing of MCMC
chains, using a modified version of the GetDist program provided with
the CosmoMC package.

\section{Results and discussion} \label{sec:results}

A first point we want to remark is that including the BAO prior
in the likelihood analysis causes just minor shifts on the limits, as
shown in Table \ref{tab:limits}, and does absolutely not conflict with
previous findings.

\begin{table}[htb]
\begin{center}
\begin{tabular}{lcc}
\hline \hline
Datasets & $\Mnu$ (eV) &$\beta$\\
\hline
WMAP5                                     				&
1.70 (1.52) & 0.136 $\pm$ 0.085 (0.098 $\pm$ 0.069) \\
WMAP5++  (without BAO)                                 	& 1.17 (1.13)& 0.104
$\pm$ 0.047 (0.100 $\pm$ 0.043) \\
WMAP5++      							& 1.19
(1.19) & 0.094 $\pm$ 0.041 (0.092 $\pm$ 0.042) \\
\end{tabular}
\caption{We report 95\% marginalized upper limits on $\Mnu$ (first
column) as well as mean value and 1$ \sigma$ uncertainties on $\beta$
for the SUGRA (RP) coupled models and different combinations of data
sets (second column).}
\label{tab:limits}
\end{center}
\vskip -.3truecm
\end{table}
Let us then consider the effects of adding mass priors.

\begin{figure}[htb]
\begin{center}
\includegraphics[height=4.cm,angle=0]{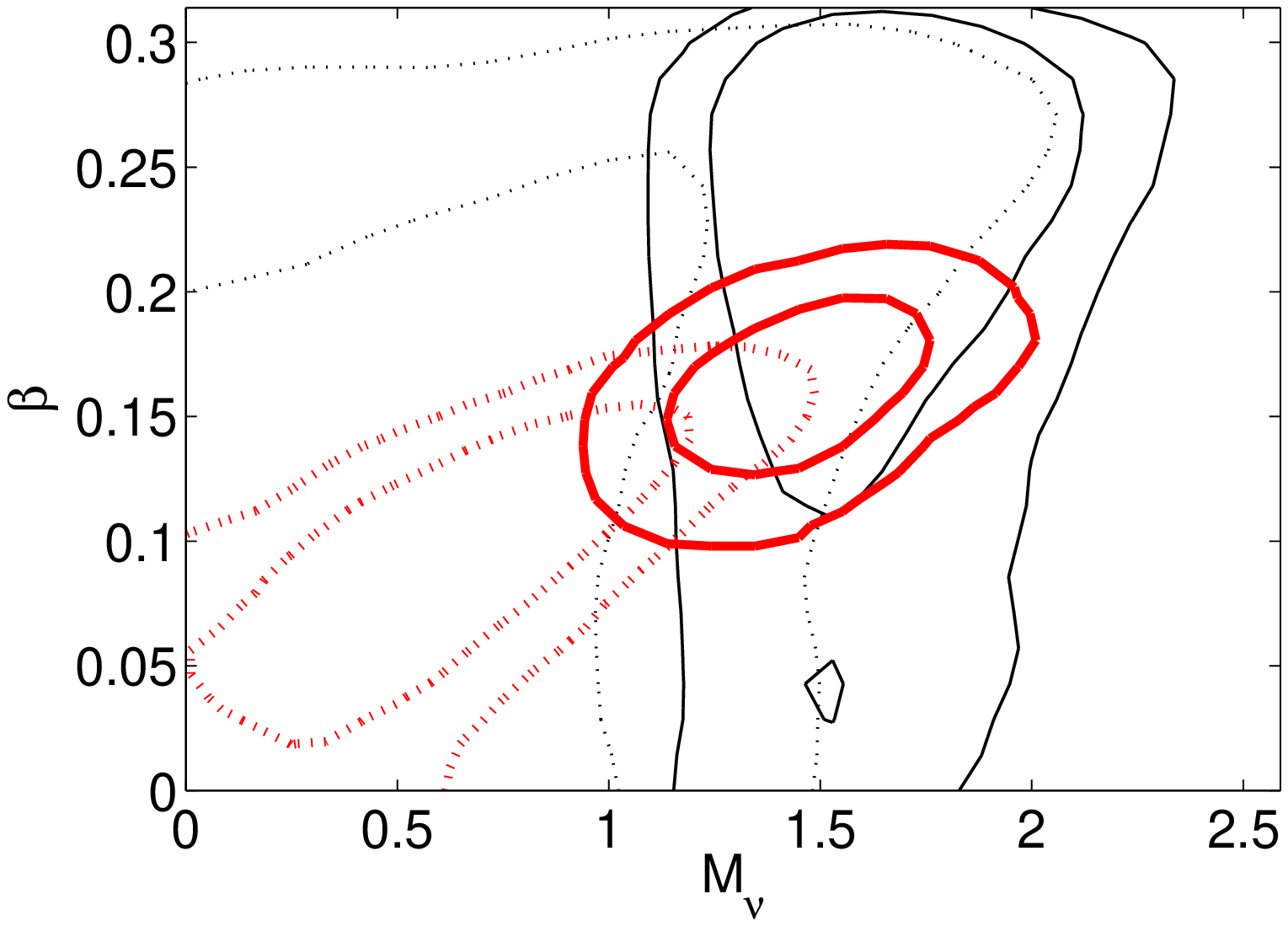}
\includegraphics[height=4.cm,angle=0]{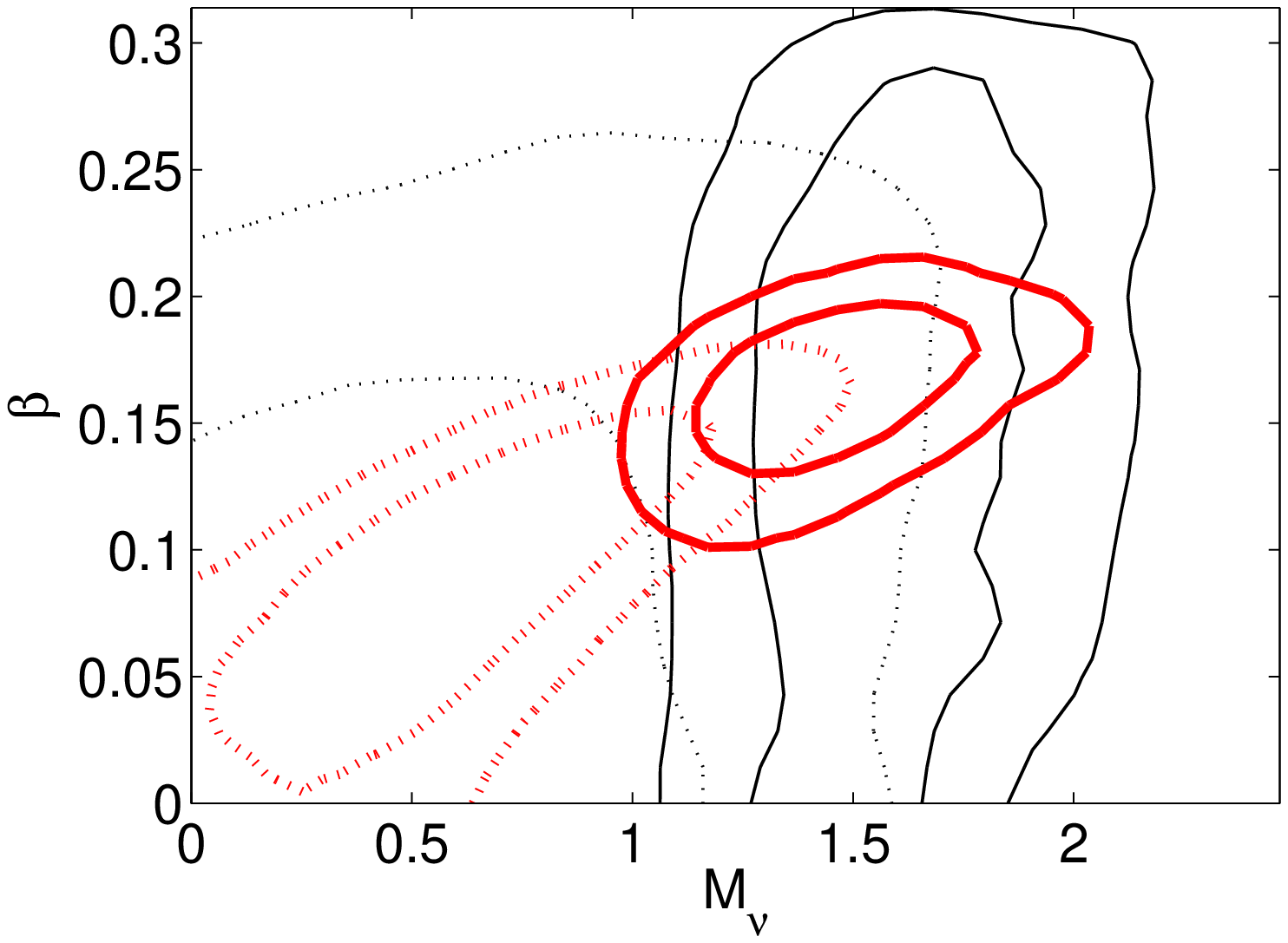}
\caption{68\% and 95\% confidence intervals in the $\Mnu-\beta$
plane. The left panel shows resulting limits when using a SUGRA
potential, while the right panel shows the limits when using a RP
potential. Black, thin lines indicate the results when using WMAP5 as
the only cosmological data set, and red, thick lines show the results
when using WMAP5++. Dotted lines are the cosmology only limits. The
resulting limits when also including the KKDC limit on $\Mnu$ are
shown with solid lines.  We notice that the
intersection between $\beta$ and $M_\nu$ allowed areas, when KKDC
controversial results are allowed or omitted, does not vanish
but is rather small.} 
\label{fig:KKDC}
\vskip +.1truecm
\end{center}
\end{figure}

 \begin{figure}[htb]
\begin{center}
\includegraphics[height=4.cm,angle=0]{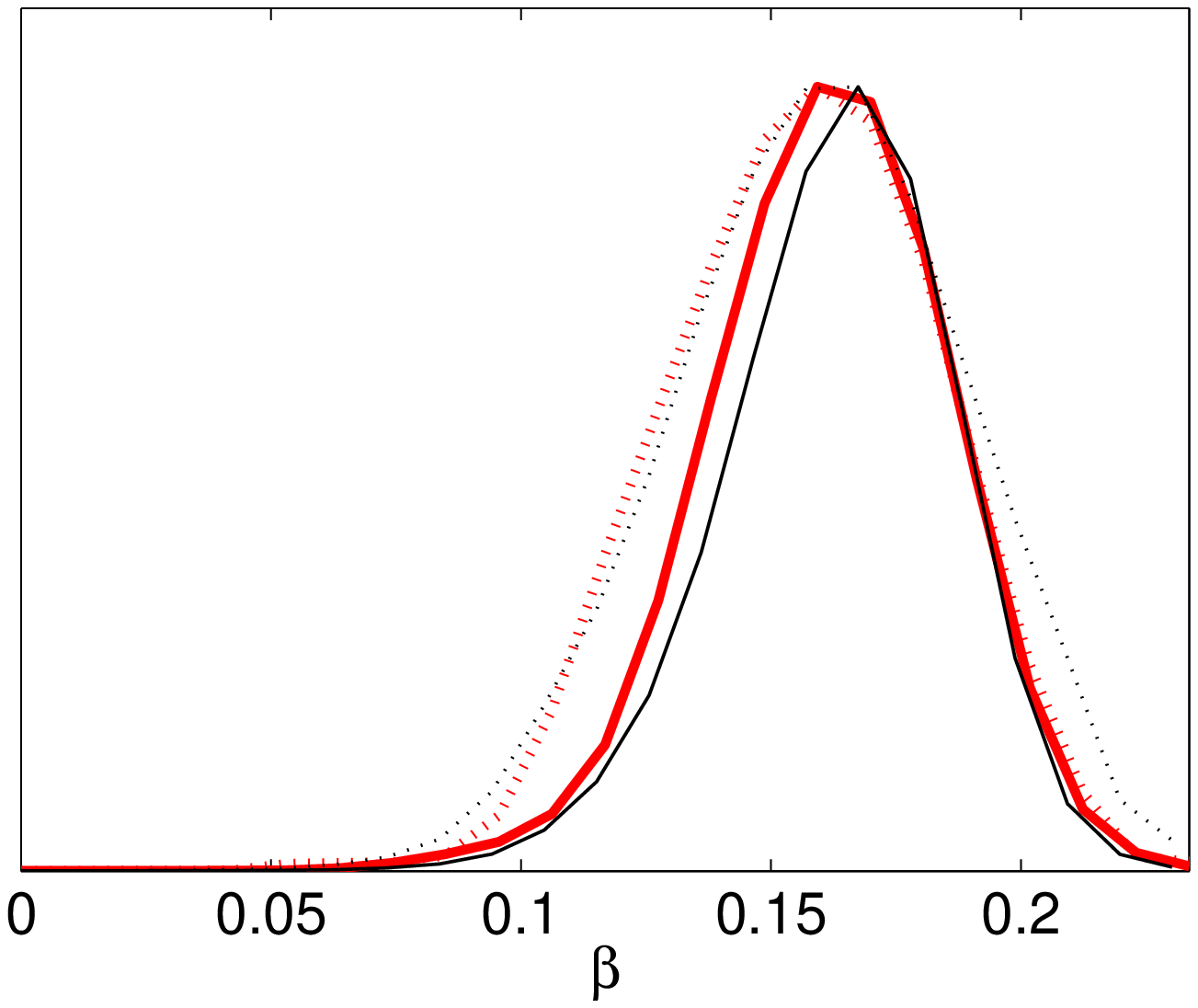}
\caption{1D likelihood distributions for $\beta$, using WMAP5++ sets
and the KKDC prior on $M_\nu$. Thick, red lines corresponds to the
SUGRA model, while thin black lines refers to the the RP model. Solid
lines denote marginalized likelihood. Dotted lines indicate average
likelihood of the samples in each bin.  }
\label{fig:KKDC_1D}
\vskip +.1truecm
\end{center}
\end{figure}

In Figure \ref{fig:KKDC} we show marginalized 68\% and 95\% confidence
contours in the $M_\nu - \beta$ plane with and without a KKDC prior on
$M_\nu$. The left (right) panel concerns the SUGRA (RP)
potential. Before including KKDC priors, results are similar to those
in our previous paper \cite{lavacca:2009}. In the WMAP5++ case,
however, where we now include BAO data, the curves shift slightly
farther from the 0-0 case, bringing the 0--0 point well outside
1$\sigma$ contours for both RP and SUGRA cases. However, without
additional priors on $\Mnu$, the 0-0 model is not excluded by
cosmological data alone.

When we add the KKDC prior on $M_\nu$, we have a significant offset
from the 0-0 model: $\beta = 0$ is excluded with $7.3~ \sigma$ ($8.0
\sigma$)in the SUGRA (RP) case. This conclusion relies on the use of
WMAP5++; CMB spectra are not sufficiently affected by $M_\nu$, to allow for a
statistical detection of $\beta$ alone, even when the KKDC prior is
imposed. This is not unexpected, as CMB data primarily probes the
Universe at high $z$, before DE became important.

Figure \ref{fig:KKDC_1D} shows 1D likelihood distributions for both
SUGRA and RP models, when using WMAP5++ and the KKDC prior. Here we
also plotted the mean likelihood for each bin, in addition to the
marginalized likelihood. In ref. \cite{lavacca:2009} a problem that
was discussed was the discrepancy between these two distributions for
the $\beta$ parameter, which was caused by a very non-gaussian
correlation between the $\log_{10} \Lambda$ and $\beta$ parameters,
especially for small values of $\beta$. From Figure \ref{fig:KKDC_1D}
we see that this problem is not very pronounced here. When $\beta$ is
forced to high values by the KKDC prior, the two probability measures
correspond quite well to each other, even though we still see some
minor shifts.

\begin{figure}[htb]
\begin{center}
\includegraphics[height=4.cm,angle=0]{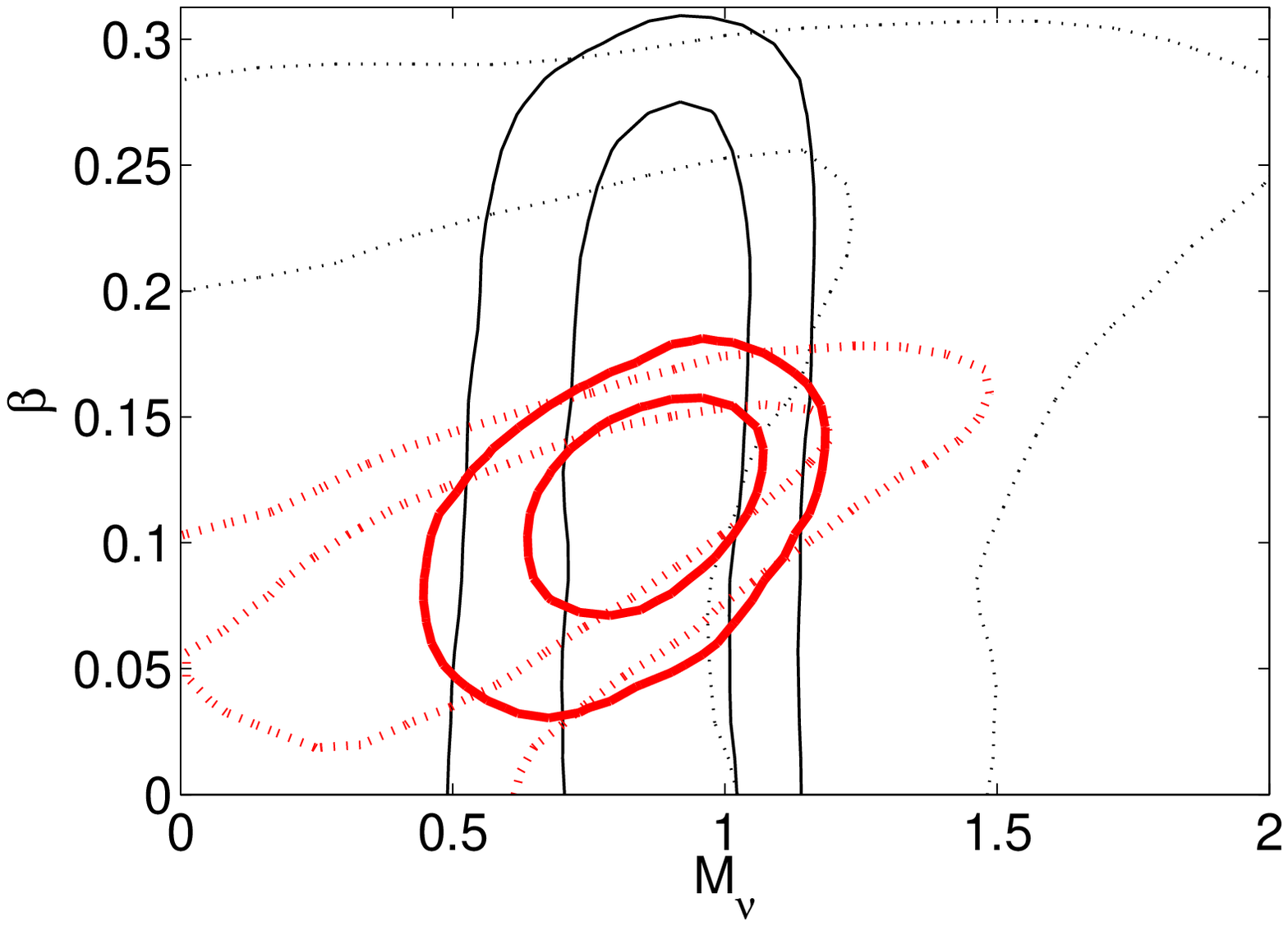}
\hskip .5truecm
\includegraphics[height=4.cm,angle=0]{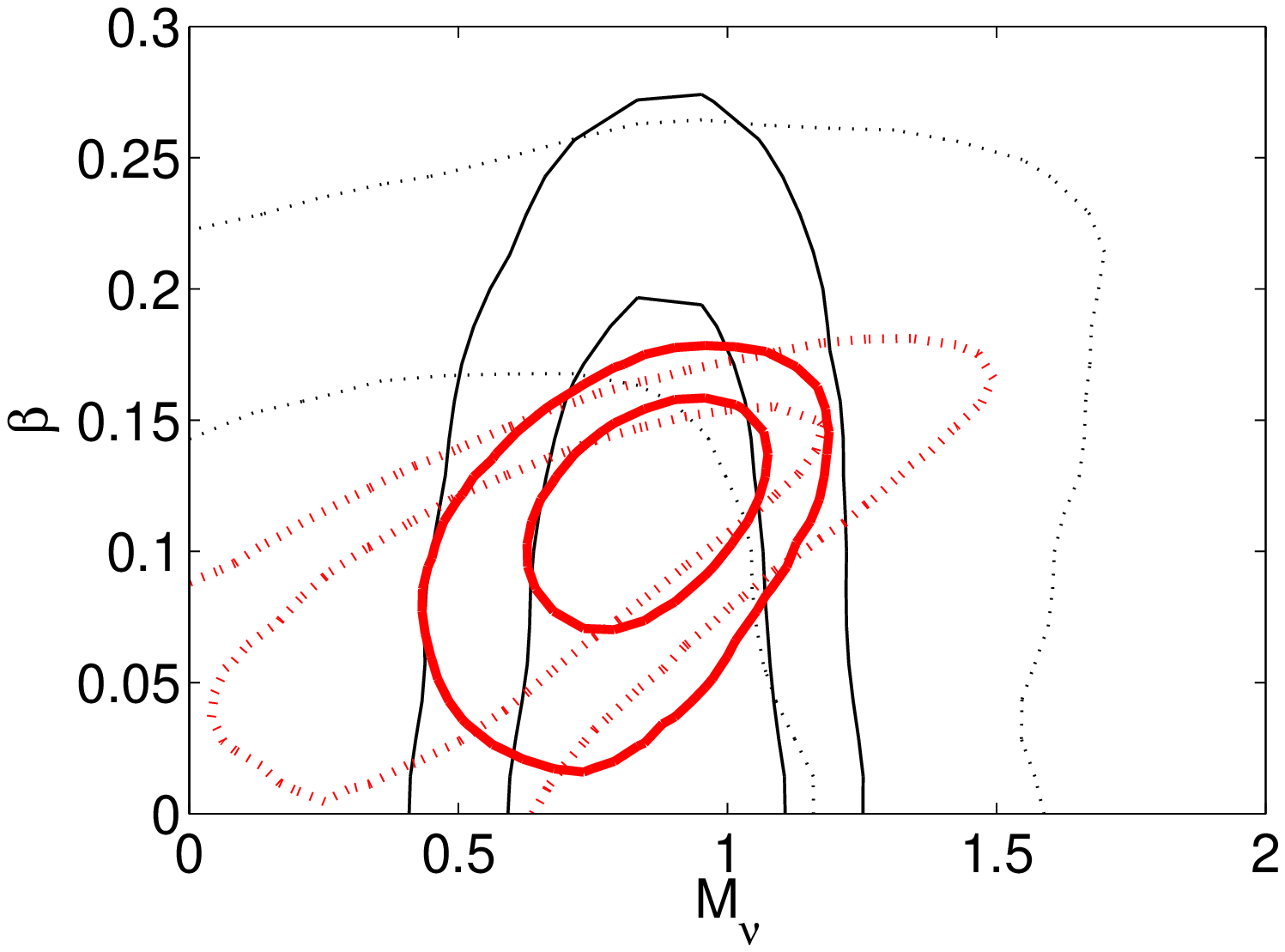}
\caption{The same as Figure \ref{fig:KKDC}, but using a KATRIN prior
with a fiducial neutrino mass of $m_\beta = 0.3$eV ($M_{\nu} = 0.9$eV)
instead of the KKDC prior. Let us also point out that, at variance from
the KKDC case, the KATRIN prior is fully consistent with cosmological
constraints, essentially leading to a restriction of the allowed area in the
$M_\nu$--$\beta$ plane.}
\label{fig:KATRIN}
\vskip +.1truecm
\end{center}
\end{figure}

In Figure \ref{fig:KATRIN} the 2D marginalized likelihood contours are
shown when the KKDC prior is replaced by the prospected prior from
KATRIN with a fiducial neutrino mass of $M_\nu = 0.9$eV. Although the
best-fit $M_\nu$ here is smaller than in the KKDC case, we still get a
significant preference for a non-zero $\beta$ when using WMAP5++. For
the SUGRA potential, $\beta=0$ gets excluded with $3.9 \sigma$
significance, while the corresponding number is $3.6 \sigma$ for the
RP potential.

The controversial results of the HM experiment, as well as the
possibility that KATRIN detects a $\nu$ mass value, had already
triggered a discussion on the way to soften cosmological constraints.
The best option put forward, up to now, was perhaps the possibility of
allowing the state parameter of DE, $w$, to delve into the {\it
phantom} regime.

\begin{figure}[htb]
\begin{center}
\includegraphics[height=6.cm,angle=0]{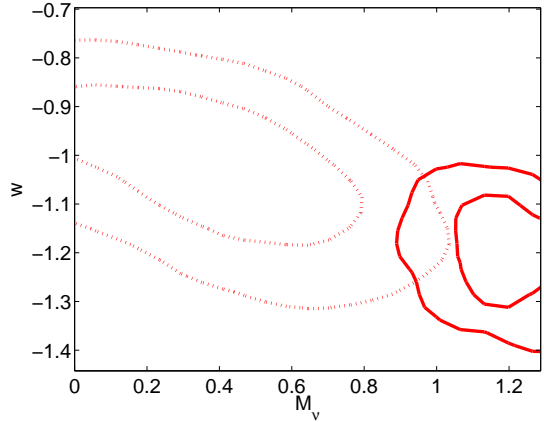}
\caption{Likelihood distribution with or without the KKDC prior
(solid or dotted lines, respectively, indicating 1 and 2 $\sigma$'s),
when a DE state parameter $w < -1$ is allowed and considering all
cosmological data. Notice that the high likelihood areas, with or without the
KKDC prior, have just a minor intersection at the 2--$\sigma$
level. Let us recall that, with the coupling option, there is a
(small) intersection even at 1--$\sigma$.}
\label{KEF}
\vskip +.1truecm
\end{center}
\end{figure}

Before concluding this Section it is then worth comparing our results
with those obtainable if we just allow for $w < -1$, an option already
deepened by \cite{KE}. Figure \ref{KEF} shows the likelihood
distributions on the $M_\nu$--$w$ plane with or without the KKDC
constraint.
This figure confirms the findings of \cite{KE}, also when the more
limited data set considered by them (including, {\it i.e.}, just WMAP3
outputs) is replaced by the whole system of data considered in the
rest of this work.

As is known, SUGRA (or RP) uncoupled cosmologies yield no significant
likelihood improvement (or worsening) in respect to $\Lambda$CDM.
They are therefore analogous benchmarks for model likelihood
confrontation. It is then clear that, if we just allow $w$ to run in
the {\it phantom} range, we hardly gain a $\sim 2$--$\sigma$
likelihood improvement. On the contrary, if we open the coupling
option, already with $M_\nu = 0.9$~eV, we approach $\sim
4$--$\sigma$'s; passing from 2 to 4--$\sigma$ means achieving
statistical significance.

However, even letting apart statistical evaluations, direct inspection
shows that spectral distortions due to increasing $M_\nu$, both in
$C_l$ and $P(k)$, are just opposite to those due to $\beta$. On the
contrary, $w < -1$ spectral changes exhibit a different scale
dependence; in fact, their capacity to allow higher $\nu$ masses is
substantially related to their favoring greater $\Omega_m$ values.

\section{Summary and conclusions} \label{sec:summary}

We have studied the effects of coupling between a dynamical DE
component and CDM. The observational effects of such a coupling are
almost opposite to those caused by massive neutrinos, which results in
a strong degeneracy between the coupling parameter $\beta$ and the
neutrino mass $M_\nu$.

This suggests the possibility that interactions within the dark sector
have been hidden to observations, up to now, by the existence of a
significant $\nu$ mass value. The $\nu$ mass, possibly responsible for
this {\it chamaleontic} effect, lays just below the range already
inspected in $H^3$ $\beta$--decay experiments. Such a range will be
however soon explored by the KATRIN experiment, while the $\nu$ mass
detection claimed by a part of the HM team is just above the suitable
range.

In this work, such a possibility was also tested including, for the
first time, BAO measurements.

Previous analysis had already shown that a cosmology with
$\nu$ mass and coupling is statistically preferred to $\Lambda$CDM.
The inclusion of BAO data slightly increases such preference
which, however, still keeps $\cal O$$(2 \sigma)$.

Such $\beta-\Mnu$ ``degeneracy'' can be broken by an external prior on
$M_\nu$ from earth based experiments. If we assume the KKDC claim on
neutrino mass  to be correct, this then results in a 7--8
$\sigma$ detection of a non-zero $\beta$. If the upcoming KATRIN
experiments confirm a neutrino mass in the range allowed by the KKDC
experiment, this will on its own standing give a statistically
significant detection of a non-zero $\beta$.

Other options considered before, to try to reconcile cosmic data
with $\nu$ mass in such range, are far less effective.

In particular, even if the option $w < -1$ is added to $\Lambda$CDM
models, a prior $M_\nu \sim 1$~eV badly modifies the likelihood
distribution, so indicating an apparent conflict with cosmological
measurements. On the contrary, when such a prior is added to
cosmological data, within the context of $\beta \neq 0$ models,
already (slightly) favored by data in respect to $\Lambda$CDM, it just
leads to a further restriction of the allowed parameter area. In a
sense, it appears a ``welcome'' new limit, just narrowing parameter
error bars.

A detection of a non-zero $m_\beta \gtrsim 0.3$eV by KATRIN would radically renew the cosmic scenario:
the concordance cosmology, $\Lambda$CDM, would be statistically
falsified; a CDM--DE coupling would become a concrete option.

One should also keep in mind that in the coming years, with data from
the Planck satellite, the CMB measurements will improve vastly, 
further narrowing the uncertainties in the $\beta-M_\nu$ plane, as
was shown in \cite{lavacca:2008}.

It may be then worth pointing out that admitting a linear interaction
between CDM and DE, as done here, is just the next approximation to
assuming them fully decoupled.  The idea underlying this analysis is
that modeling the Dark Cosmic Sector with two independent components
is just a first step towards understanding its complex nature. After a
hypothetical confirmation of the coupling option by laboratory data,
further cosmic data might allow to go even beyond the assumption of
linear interaction, possibly providing a more detailed insight into
one of basic question of modern physics, the nature of the cosmic dark
components.

\section*{Acknowledgments}
JRK and RM acknowledge support from the Research Council of
Norway. The computations presented in this paper were carried out on
Titan, a cluster owned and maintained by the University of Oslo and
NOTUR.


\vfill

\end{document}